\documentclass[12pt,preprint]{aastex}

\usepackage{amsmath}
\usepackage{subfig}
\usepackage{rotating}

\begin{document}
%

\title{Temperature Fluctuations as a Source of Brown Dwarf Variability}

\author{Tyler D. Robinson}
\affil{NASA Ames Research Center, MS 245-3, Moffett Field, CA 94035, USA}
\email{tyler.d.robinson@nasa.gov}

\and

\author{Mark S. Marley}
\affil{NASA Ames Research Center, MS 245-3, Moffett Field, CA 94035, USA}

%
\begin{abstract}

A number of brown dwarfs are now known to be variable with observed amplitudes 
as large as 10--30\% at some wavelengths. While spatial inhomogeneities in cloud 
coverage and thickness are likely responsible for much of the observed variability, it 
is possible that some of the variations arise from atmospheric temperature fluctuations 
instead of, or in addition to, clouds. To better understand the role that thermal variability 
might play we present a case study of brown dwarf variability using a newly-developed 
one-dimensional, time-stepping model of atmospheric thermal structure.  We focus on 
the effects of thermal perturbations, intentionally simplifying the problem through omission 
of  clouds and atmospheric circulation.  Model results demonstrate that thermal 
perturbations occurring deep in the atmosphere (at pressures greater than 10~bar) of a 
model T-dwarf can be communicated to the upper atmosphere through radiative heating 
via the windows in near-infrared water opacity.  The response time depends on where in 
the atmosphere a thermal perturbation is introduced.  We show that, for certain periodic 
perturbations, the emission spectrum can have complex, time- and wavelength-dependent 
behaviors, including phase shifts in times of maximum flux observed at different 
wavelengths.  Since different wavelengths probe different levels in the atmosphere, these 
variations track a wavelength-dependent set of radiative exchanges happening between 
different atmospheric levels as a perturbation evolves in time.  We conclude that 
thermal---as well as cloud---fluctuations must be considered as possible contributors to 
the observed brown dwarf variability.
\end{abstract}


\section{Introduction}
After more than a decade and a half of surveys for brown dwarf variability we now know that 
the emergent spectrum of many L- and T- type brown dwarfs indeed varies with time 
\citep[e.g.,][]{tinney&trolley1999,bailerjones&mundt1999,artigauetal2009,radiganetal2012,buenzlietal2013}.
Broadband, near-infrared  flux variations can be as large as 10--30\%, can occur on timescales 
from 1--100~hours, and can be non-periodic 
\citep{bailerjones&mundt2001a,gelinoetal2002,artigauetal2009,radiganetal2012,gillonetal2013}.  
Spectroscopic and multi-band photometric studies have revealed complex, wavelength-dependent 
lightcurves \citep{radiganetal2012,buenzlietal2013}.  In some cases, spectra show periodic 
brightness fluctuations with wavelength-dependent phase lags, which can be as large as 
180$^{\circ}$ (i.e., shifted by half of a cycle) \citep{buenzlietal2012}.

Clouds sculpt the emergent spectra of essentially all spectral classes of brown dwarfs, 
although their impact is most notable in the L-dwarfs 
\citep{leggettetal1998,chabrieretal2000,allardetal2001,ackerman&marley2001,tsuji2002,
golimowskietal2004,knappetal2004,burrowsetal2006,stephensetal2009,morleyetal2012}. 
Given the strong evidence for the presence of clouds in brown dwarf atmospheres, and 
the ability of a continuum opacity source to limit the depth of the wavelength-dependent 
photosphere \citep{ackerman&marley2001}, it is expected that these structures play some 
role in brightness variability \citep{radiganetal2012,apaietal2013}.  Indeed thermal emission 
from the deep atmospheres of both Jupiter and Saturn is strongly modulated by cloud 
structures and Jupiter itself would show substantial variability if the disk were observed at 
5~$\mu$m in integrated light \citep{gelino&marley2000}.  However the intensity of radiation 
emitted by a planetary or brown dwarf atmosphere depends on many factors in addition to 
cloud structure.  Atmospheric temperature and composition also control the thermal emission 
and it seems prudent to also consider the role such factors might contribute to variability.

\citet{freytagetal2010} used a 2-D radiation hydrodynamics model to study atmospheric 
circulation and dust transport in M-dwarf and brown dwarf atmospheres.  This work 
highlighted the importance of gravity waves and dust convection to maintaining clouds in 
brown dwarf atmospheres.  They found that gravity waves are expected to be ubiquitous 
above the radiative-convective boundary and likely play an important role in cloud 
development and evolution.  More recently, global 3-D, cloud-free models were used by 
\citet{showman&kaspi2013} to study large-scale flows and convection in the interiors 
and deep atmospheres of brown dwarfs.  These models revealed that convection is 
strongly influenced by the relatively fast rotation rates of brown dwarfs and that thermal 
variations of order several Kelvin may be expected at the top of the model convective 
zone.  This work also discussed a stratospheric circulation, driven by the interaction 
between atmospheric waves generated at the top of the convective zone and  the mean 
stratospheric flow, that could lead to large ($\sim$50~K) temperature variations in 
the upper atmospheres of brown dwarfs.  However, emission spectra were not computed, 
and so the influence of these variations on the brown dwarfs' spectra remain unclear.

Here we provide a case study of the impact of atmospheric temperature fluctuations on 
the emission spectrum of a brown dwarf.  We intentionally simplify the problem by 
neglecting clouds as well as chemical evolution, thus allowing us to explore the behaviors, 
timescales, and related spectral variability due to temperature fluctuations alone.  Using a 
new 1-D, time-stepping radiative convective model for brown dwarfs, we first investigate the 
heating of an atmosphere due to an extended thermal pulse from deep within the convective 
zone.  We then explore variability in the emission spectrum due to time-varying thermal 
fluctuations, introducing these perturbations at different atmospheric levels, and highlighting 
circumstances where the model can reproduce wavelength-dependent phase lags.

\section{Model Description}
We adapt a well-validated thermal structure model of brown dwarfs 
\citep{marleyetal1996,marleyetal2002} to permit realistic time-stepping, thus permitting studies 
of the time-dependent evolution of atmospheric thermal pulses.  We take the atmosphere to 
be in hydrostatic equilibrium.  The heating rate, $Q$, throughout the atmospheric profile is 
given by
\begin{equation}
Q = \frac{dT}{dt} = \frac{g}{c_{p}}\frac{dF_{\text{net}}}{dp} \ ,
\label{eqn:heating}
\end{equation}
where $T$ is temperature, $t$ is time, $g$ is the acceleration due to gravity, which we take to 
be constant over the small range of altitudes encompassed by the atmosphere, $c_{p}$ is the 
temperature- and pressure-dependent specific heat capacity, $F_{\text{net}}$ is the net energy 
flux, carried by both radiation and convection, and $p$ is pressure.  Values for the specific heat 
are taken from the equation of state models of \citet{saumonetal1995}.  

Gas opacities in our model are computed according to \citet{freedmanetal2008}, and chemical 
abundances of all radiatively active species are provided by the equilibrium models of 
\citet{lodders2004} and \citet{lodders&fegley2006}.  For a given pressure-temperature ($p$-$T$) 
profile and set of gas composition profiles, the upwelling and downwelling thermal fluxes are 
computed for 180 wavelength bins, spanning 0.4--325~$\mu$m.  Within each spectral interval, 
the two-stream source function technique \citep{toonetal1989}\footnote{Note that the ``two-stream'' 
description applies to an intermediate approximation used to speed the cloud scattering 
computation and that this method is not a classical two stream calculation. Rather the radiative 
transport is computed with five to ten angular beams as necessary.} is used to solve the  
equation of radiative transfer within discrete spectral bins using eight-term, 
correlated-$k$ coefficients.  

The convective heat flux, $F_{c}$, is computed according to mixing length theory 
\citep{vitense1953,gierasch&goody1968}, and is given as 
\begin{equation}
F_{c} = - \rho c_{p} K_{H} \left( \frac{dT}{dz} + \Gamma_{\text{ad}} \right) \ ,
\end{equation}
where $\rho$ is the mass density, $K_H$ is the eddy diffusivity for heat, $z$ is altitude, and 
$\Gamma_{\text{ad}}=g/c_p$ is the adiabatic lapse rate.  The eddy diffusivity vanishes when 
the temperature profile is stable against convection, and is given by
\begin{equation}
K_{H} = 
  \begin{cases}
    l^{2} \left[ \frac{g}{T} \left( \frac{dT}{dz} + \Gamma_{\text{ad}} \right) \right]^{1/2}, & 
    \frac{dT}{dz} > - \Gamma_{\text{ad}}  \\ 0, &  \frac{dT}{dz}  \leq - \Gamma_{\text{ad}}
  \end{cases}
\end{equation}
where $l$ is the mixing length, which we set equal to the pressure scale height, $H$, with
\begin{equation}
l = H = \frac{k_{B}T}{\mu g} \ ,
\end{equation}
where $k_{B}$ is the Boltzmann constant, and $\mu$ is the atmospheric mean molecular 
mass.  We tested sensitivity to our mixing length parameterization by running simulations 
with $l=2H$ and $l=\frac{1}{2}H$, and found negligible differences in model outputs.

Time-stepping in our model proceeds in a straightforward, linear fashion.  A time increment, 
$\Delta t$, is selected that is short enough to maintain stability in the convective region of 
the atmosphere.  Typically, $\Delta t$ of order 0.1--1~s suffices, which is about 1/10 of the 
convective timescale for a single model layer.  Given a $p$-$T$ profile at time $t$, the 
temperature at level $i$ is updated according to
\begin{equation}
T_{i}\left(t+\Delta t\right) = T_{i}\left(t\right) + Q_{i}\Delta t \ ,
\end{equation}
where $Q_{i}$ is the heating rate at model pressure level $i$.  Our pressure grid contains 
68 levels between 90~bar and $3\times 10^{-4}$~bar which are, roughly, equally spaced in
log-pressure.  The wall-clock time of each model time-step is much less than one second, 
due to the computational efficiency of our radiation and convection schemes.  We verified that 
the time-stepping model's equilibrium $p$-$T$ curve and depth of the convective zone agreed 
with previous versions of our model, which found equilibrium solutions using a 
Newton-Raphson technique.

In this case study, we use a model T dwarf with an effective temperature of 900~K and a surface 
gravity of $10^{5}$~cm~s$^{-2}$.  This set of parameters was previously found to produce 
steady-state models that reproduce observations of a variable T6.5 brown dwarf, 
2MASS~J22282889--431026 \citep{buenzlietal2012}.  Note that the best-fit models for 
this case included thin clouds of Na$_2$S, MnS, and Cr, which, as mentioned earlier, we 
intentionally omit.  

\begin{figure}
  \centering
  \includegraphics[scale=0.9]{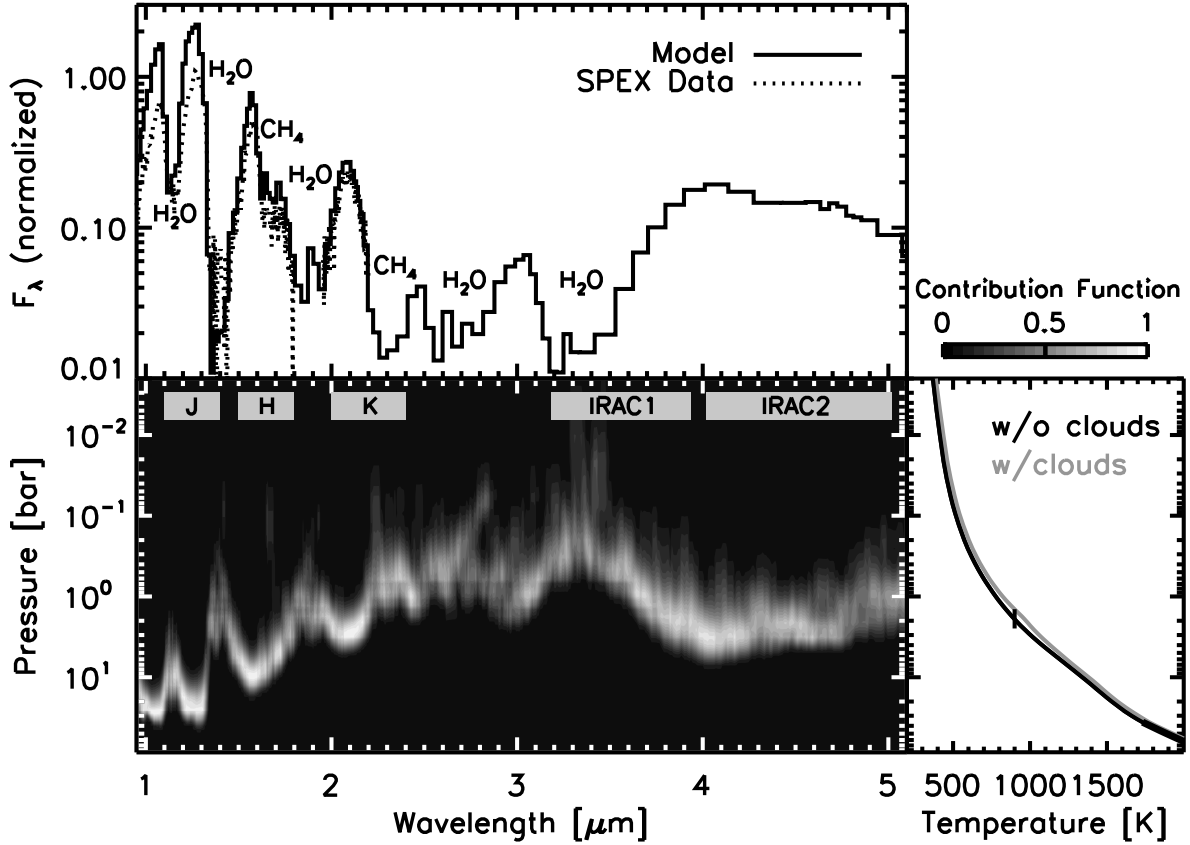}

  \caption{Top-of-atmosphere spectrum, contribution functions, and thermal structure for 
                the equilibrium state of the brown dwarf used in this case study ($T_{\text{eff}}=900$~K, 
                $g=10^{5}$~cm~s$^{-2}$).  Key absorption features and relevant near-infrared bandpasses 
                are indicated.  Also shown are SPEX observations of 2MASS J22282889--431026 
                \citep{burgasseretal2004}, which has been shown to be variable \citep{buenzlietal2012}.  
                The model is brighter than the SPEX data at continuum wavelengths due to our omission 
                of clouds.  The contribution functions indicate the range of pressures that contribute flux 
                to the top-of-atmosphere spectrum at a given wavelength (model levels are apparent in 
                the shading).  The black $p$-$T$ profile is for the standard, cloud-free case used in this 
                study.  The convective portion of the atmosphere is shown by a thickened line, and the 
                vertical mark indicates where the atmosphere temperature equals the effective temperature.  
                Also shown is a $p$-$T$ profile (in grey) from a cloudy model\citep{morleyetal2012} which 
                was found to reproduce, on average, a spectrum of the aforementioned 2MASS object in 
                \citet{buenzlietal2012}. }
  \label{fig:confuncs_spec_temp}
\end{figure}

Figure~\ref{fig:confuncs_spec_temp} shows the modeled emission spectrum, wavelength- 
and pressure-dependent normalized contribution functions \citep[][p. 131]{haneletal1992}, 
and equilibrium $p$-$T$ profile for this object, and also shows the observed SPEX spectrum 
\citep{burgasseretal2004}.  The contribution functions show the range of pressures 
that contribute information to the top-of-atmosphere spectrum (i.e., the wavelength-dependent 
specific flux exiting the atmosphere to space), and the pressure at the base of the atmospheric 
simulations was chosen to extend below the deepest pressures probed by the spectra (which 
occurs primarily at visible wavelengths).  Relevant near-infrared bandpasses are also shown, 
which, for example, demonstrate that the IRAC1 filter is sensitive to lower pressures (higher 
altitudes) than is J-band.  Small discrepancies at continuum wavelengths between the modeled 
and observed spectra are due to our omission of clouds.  For comparison, 
Figure~\ref{fig:confuncs_spec_temp} also shows a $p$-$T$ profile from a cloudy model 
\citep{morleyetal2012} that was shown to reproduce a time-averaged spectrum of the 
aforementioned 2MASS object in \citet{buenzlietal2012}.  As was shown in  
\citet{buenzlietal2012}, models of this object that include clouds place these relatively deep in 
the atmosphere, with several different cloud decks forming below 2 bar.  Our contribution 
functions demonstrate that there are wavelength regions below 1.7~$\mu$m (between the 
strong water bands) which are sensitive to these pressures.  As a result, a model spectrum 
that includes clouds has less flux at these continuum wavelengths.

\section{Results}
We proceed with two key investigations in our case study.  First, we use our time-stepping 
model to study the monotonic heating of a brown dwarf atmosphere due to an energy 
source at depth.  The focus of this investigation is to understand how the deep atmosphere 
communicates energy to the upper, radiatively-dominated levels of the atmosphere.  
Second, we introduce periodic perturbations at different levels of the atmosphere.  We do 
not specify the mechanism driving these fluctuations, but energy sources could include 
dynamics, atmospheric wave breaking \citep[e.g.,][]{youngetal1997}, cloud radiative and latent 
heating affects, and/or other processes.  As different wavelengths in the top-of-atmosphere 
spectrum probe different pressures, we show how the time-dependent perturbations lead to 
complex behavior in the emergent spectrum.

\subsection{Time-Dependent Response to Heating at Depth}
We can gain insight into the time-dependent response of a brown dwarf atmosphere 
to a perturbation by varying the bottom boundary condition, which is the internal heat 
flux.  This flux is equal to $\sigma T_{\text{eff}}^{4}$, where sigma is the Stefan-Boltzmann 
constant, and $T_{\text{eff}}$ is the dwarf's effective temperature.  For our 
$T_{\text{eff}}=900$~K body, this flux is nominally $3.72 \times 10^{4}$~W~m$^{-2}$.  We 
study the response of the atmosphere, initially in equilibrium, to a 10\% increase 
in this flux, to $4.09 \times 10^{4}$~W~m$^{-2}$ (or $T_{\text{eff}}=922$~K).

\begin{figure}
  \centering
  \includegraphics[scale=0.9]{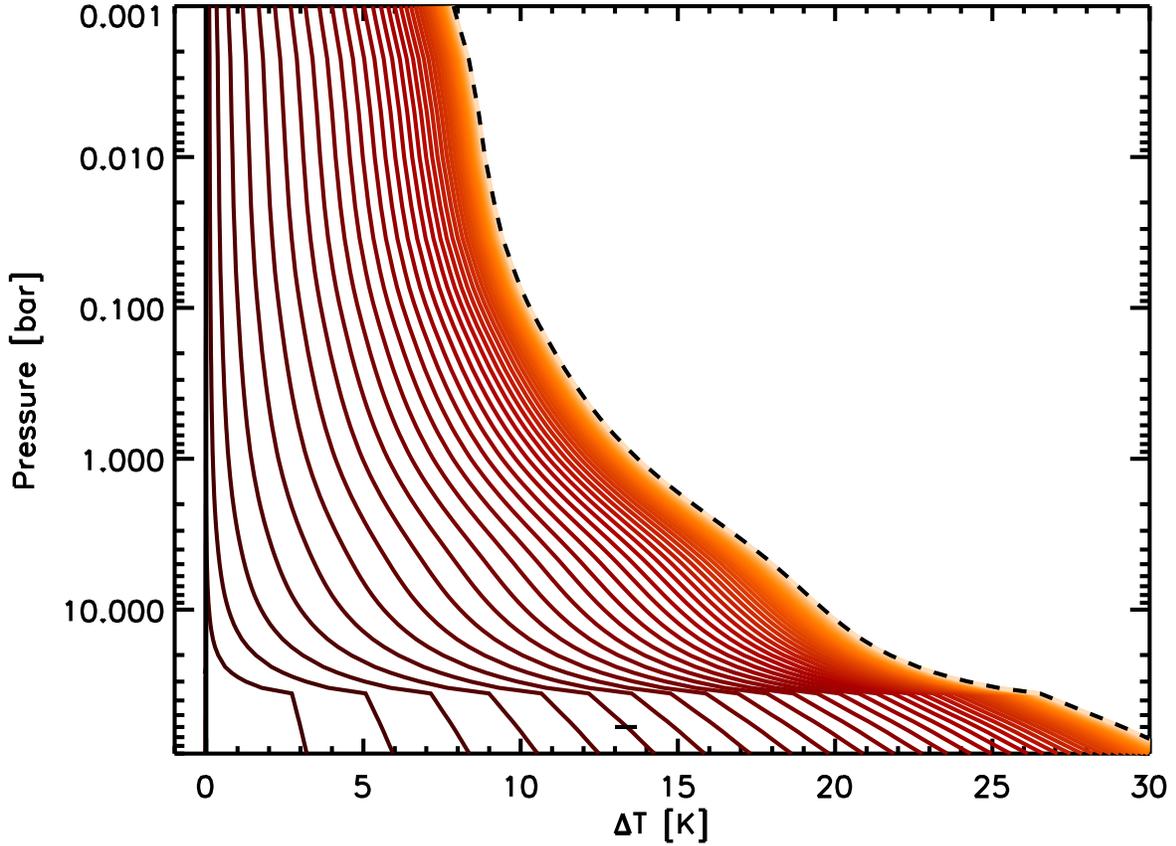}
  
  \caption{Time-dependent heating of model atmosphere in response to an increase in the 
                internal heat flux, which was increased by 10\% from 
                $3.72 \times 10^{4}$~W~m$^{-2}$ to $4.09 \times 10^{4}$~W~m$^{-2}$.  Temperatures 
                are relative to the equilibrium state shown in Figure~\ref{fig:confuncs_spec_temp}.  
                Curves are separated by 20~hours, brighter hues are for later times, and a total elapsed 
                time of 1000~hours is shown.  Note the different behaviors in the convective region 
                and the radiative region, where the radiative-convective boundary is at 40~bar.  The 
                dashed line shows the new equilibrium state, where temperatures are 1--2\% greater 
                than their initial values.  The horizontal tick mark indicates the curve for which radiative 
                heating rates are shown in Figure~\ref{fig:heating_rates}.}
  \label{fig:heating_temps}
\end{figure}

Figure~\ref{fig:heating_temps} shows the time-dependent evolution of atmospheric 
temperatures, relative to the equilibrium profile shown in 
Figure~\ref{fig:confuncs_spec_temp}.  A curve is shown for every 20~hours of 
evolution, and brighter hues show later times.  The convective portion of the 
atmosphere, which is below 40~bar, heats nearly uniformly,  while the upper portions 
of the atmosphere lag behind, which is because the convective timescale 
($\sim$0.1--1~hours) is short compared to the radiative timescale 
($\sim$10--100~hours).  Note that the new equilibrium temperatures, shown by the 
dashed curve, are about 1--2\% larger than the initial equilibrium state.  Also, this 
figure demonstrates that the characteristic timescale at which the atmosphere 
responds to the perturbation at depth is $\sim100$ hours.

\begin{figure}
  \centering
  \includegraphics[scale=0.9]{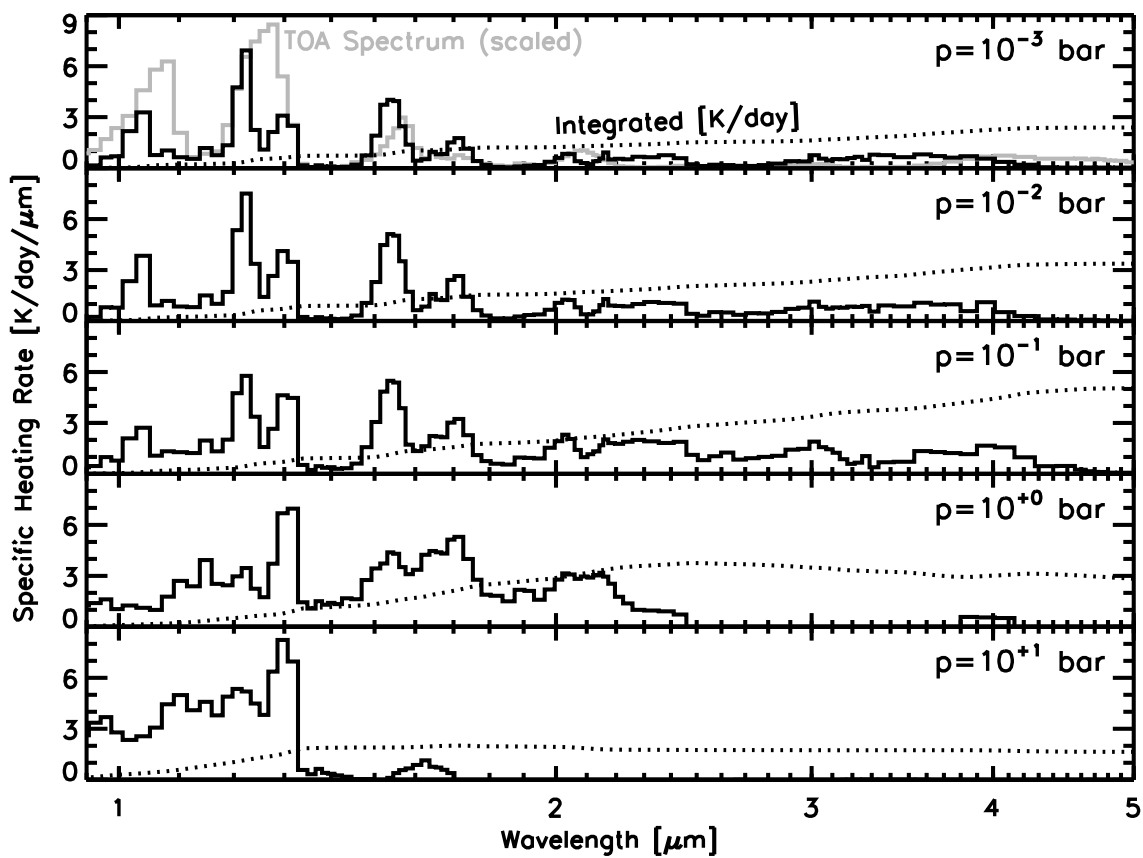}
  
  \caption{Wavelength dependent specific heating rates at 120~hours into the simulation shown in 
                Figure~\ref{fig:heating_temps} (see marked curve in this figure), which is when the radiative 
                heating rates in the upper atmosphere are at their largest.  The rates shown here have the 
                initial equilibrium rates subtracted off, and, thus, represent a net heating above the initial 
                state.  Different sub-figures are for different pressure levels in the atmosphere, which 
                are indicated in the upper-right.  The dotted curve is the running total of the integrated 
                heating rate, which uses the same vertical axis scale.  A scaled version of the 
                top-of-atmosphere spectrum is shown in the top sub-panel (in light grey).}
  \label{fig:heating_rates}
\end{figure}

Despite the perturbation being introduced at the base of the atmosphere, at 
90~bar, we see that the upper atmosphere is already beginning to heat by 
several tens of hours into the simulation, and achieves its maximum heating rate 
at about 120 hours into the simulation.  To understand this behavior, we 
investigated the wavelength- and pressure-dependent radiative heating rates at 
120 hours into the run.  These are shown in Figure~\ref{fig:heating_rates}.  Here, 
different vertical sub-figures are for different pressures in the atmosphere, at $10^{-3}$, 
$10^{-2}$, $10^{-1}$, $1$, and $10$ bar.  The horizontal axis is wavelength, and 
the vertical axis is the specific heating rate (in K~day$^{-1}$~$\mu$m$^{-1}$).  The 
dotted curve is the integrated heating rate, (in K~day$^{-1}$).  For convenience, a 
scaled version of the top-of-atmosphere spectrum is shown in light grey in the top 
sub-figure.

Investigation of the radiative heating rates in Figure~\ref{fig:heating_rates} shows 
that the upper-most portions of the atmosphere are responding to increases in radiative 
fluxes at short wavelengths (below about 2~$\mu$m).  Heating at these levels occurs 
primarily in the wings of water vapor absorption bands.  The opacity at band centers 
is too large for the upper atmosphere to ``see" flux from the deep atmosphere, and 
there is not enough opacity in window regions at low pressures for significant flux to 
be absorbed, which would drive heating at these levels.

\subsection{Periodic Perturbations at Different Atmospheric Pressures}
To investigate periodic variability, we artificially introduce a sinusoidally-varying 
heating rate to our model atmosphere at different pressures.  This perturbation is 
in addition to the heating/cooling due to gradients in the net radiative and convective 
fluxes that occur as the atmosphere evolves through time.  Thus, 
Equation~\ref{eqn:heating} is now
\begin{equation}
\frac{dT}{dt} = \frac{g}{c_{p}}\frac{dF_{\text{net}}}{dp} + Q_{f} \sin \left( \frac{2\pi t}{P} \right)\delta\left(p-p_{f}\right) \ ,
\end{equation}
where $Q_{f}$ is the amplitude of the fluctuation (in degrees Kelvin per unit time), 
$P$ is the fluctuation period, $\delta$ is the Dirac delta function, and $p_{f}$ is the pressure 
where the fluctuation is introduced.  In practice, since model levels are discretized, the heating 
perturbation is introduced over a small range of model pressures.  We choose $p_{f}$ as either the 
base of the atmosphere (near 100 bar), at 10 bar, or at 1 bar, and study the quasi-steady-state 
behavior over the course of entire fluctuation cycle.  The timescales we investigate are 10, 50, 100, 
and 500~hours, and $Q_{f}$ is selected for each run so that the temperature at the 
perturbed level varies by 1\% over a cycle (i.e., 20~K variation at the base, 13~K at 
10~bar, and 8~K at 1 bar).  As mentioned earlier, we do not specify the mechanism that 
drives these fluctuations, but possible sources include dynamics, wave breaking, cloud effects, 
and/or other processes.

\begin{sidewaysfigure}
  \captionsetup[subfigure]{labelformat=empty}
  \centering
  \begin{tabular}{cccc}
  \subfloat[]{\includegraphics[width=2.25in]{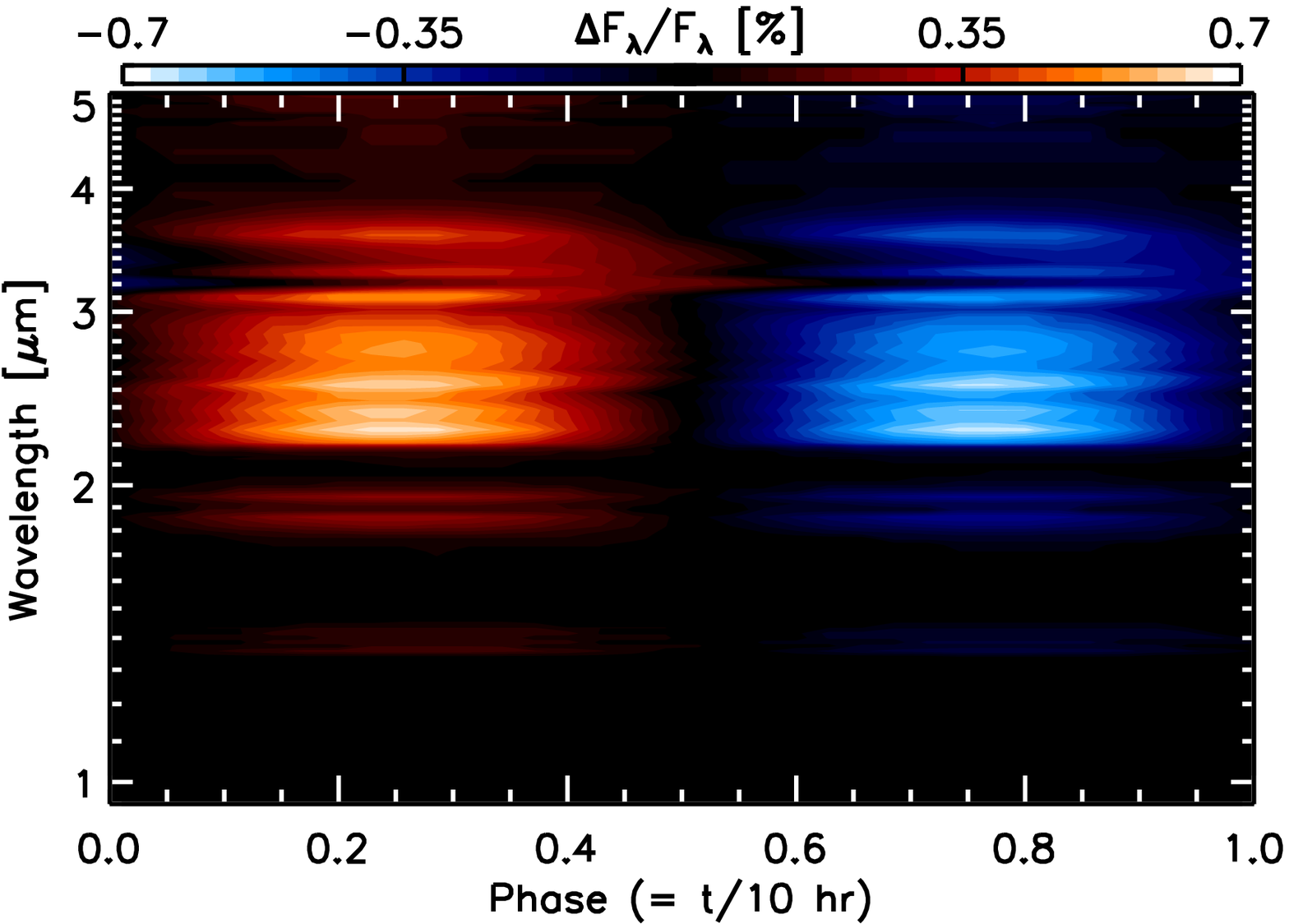}} & 
  \subfloat[]{\includegraphics[width=2.25in]{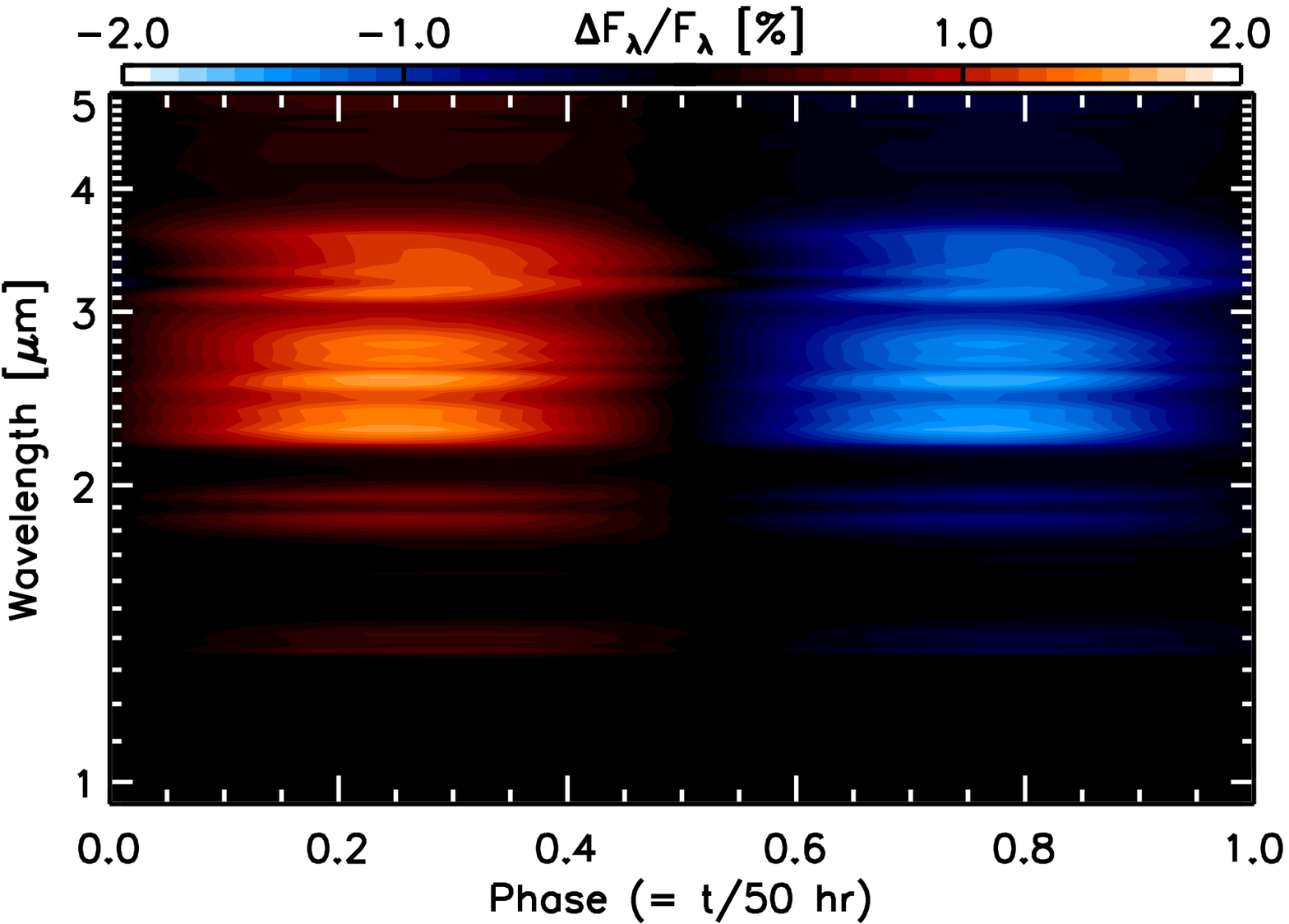}} & 
  \subfloat[]{\includegraphics[width=2.25in]{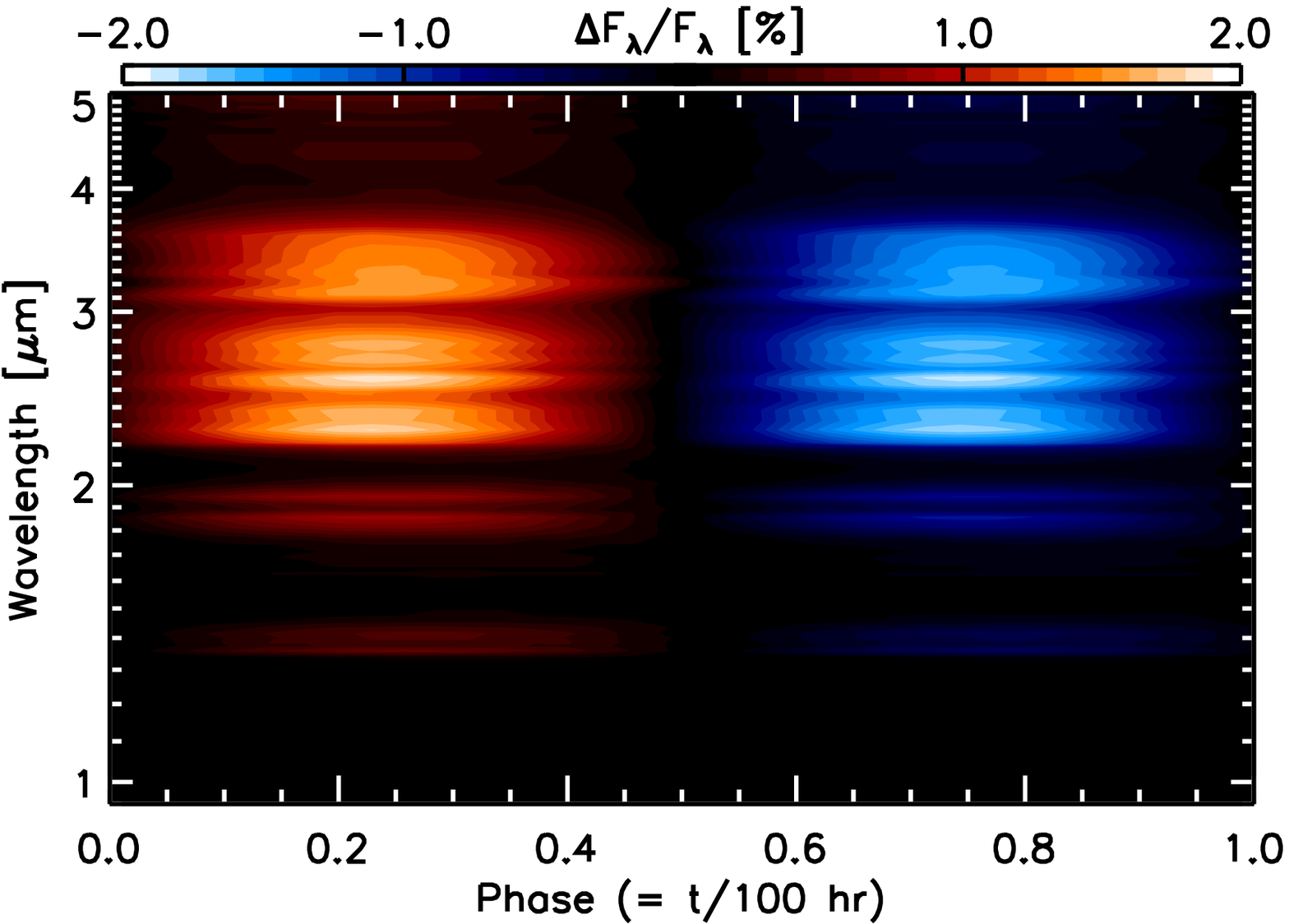}} &
  \subfloat[]{\includegraphics[width=2.25in]{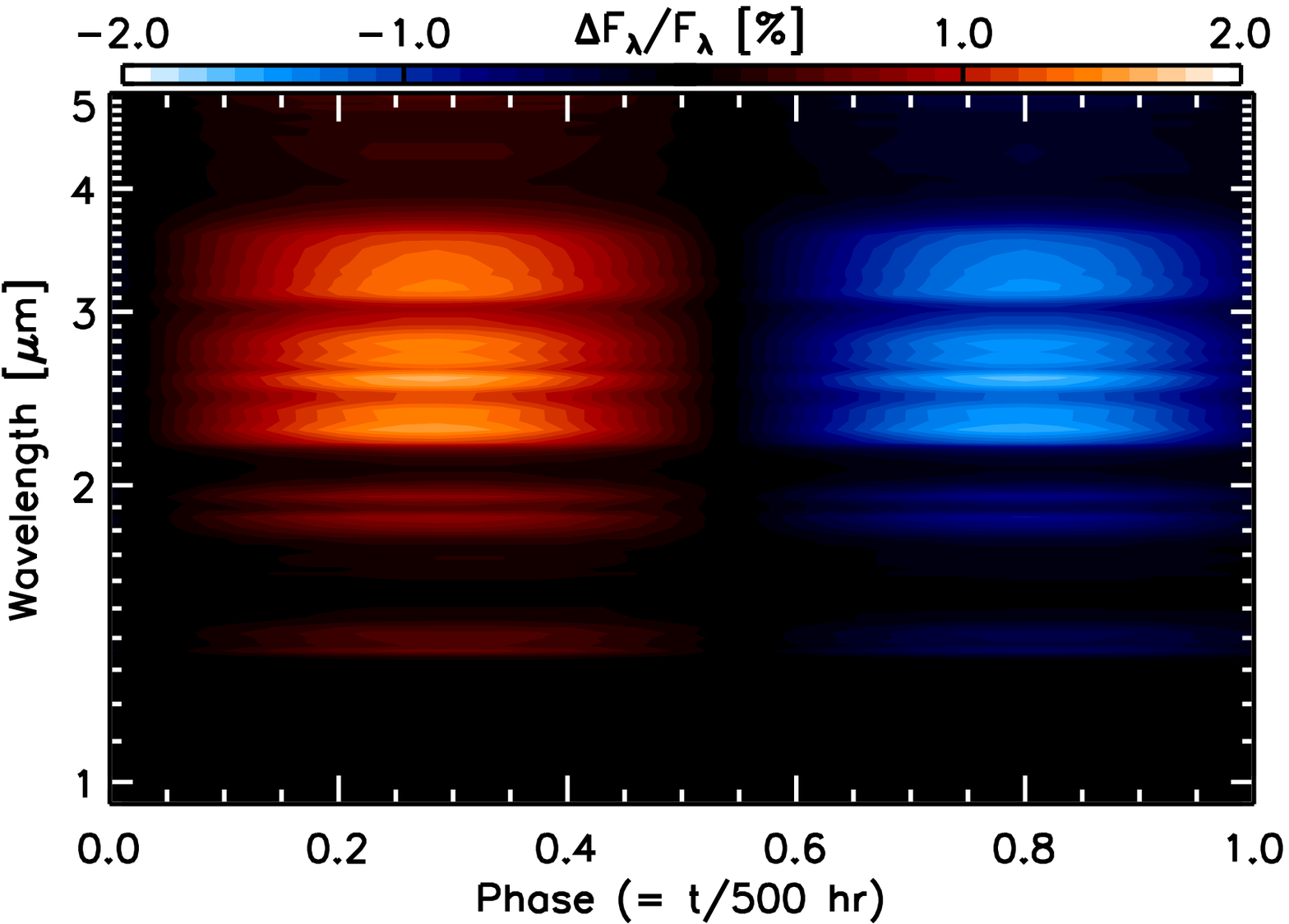}}
  \\
  \subfloat[]{\includegraphics[width=2.25in]{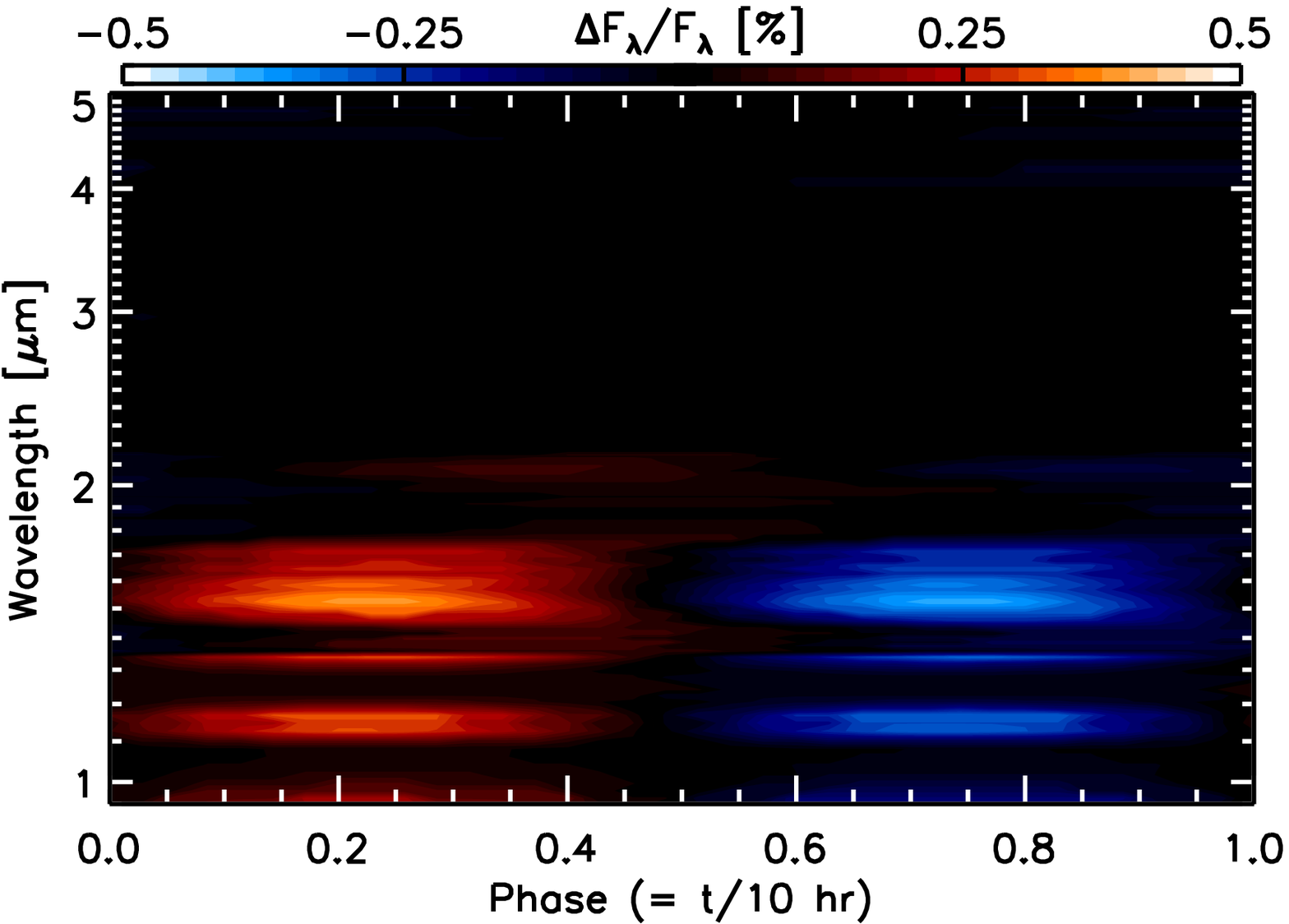}} & 
  \subfloat[]{\includegraphics[width=2.25in]{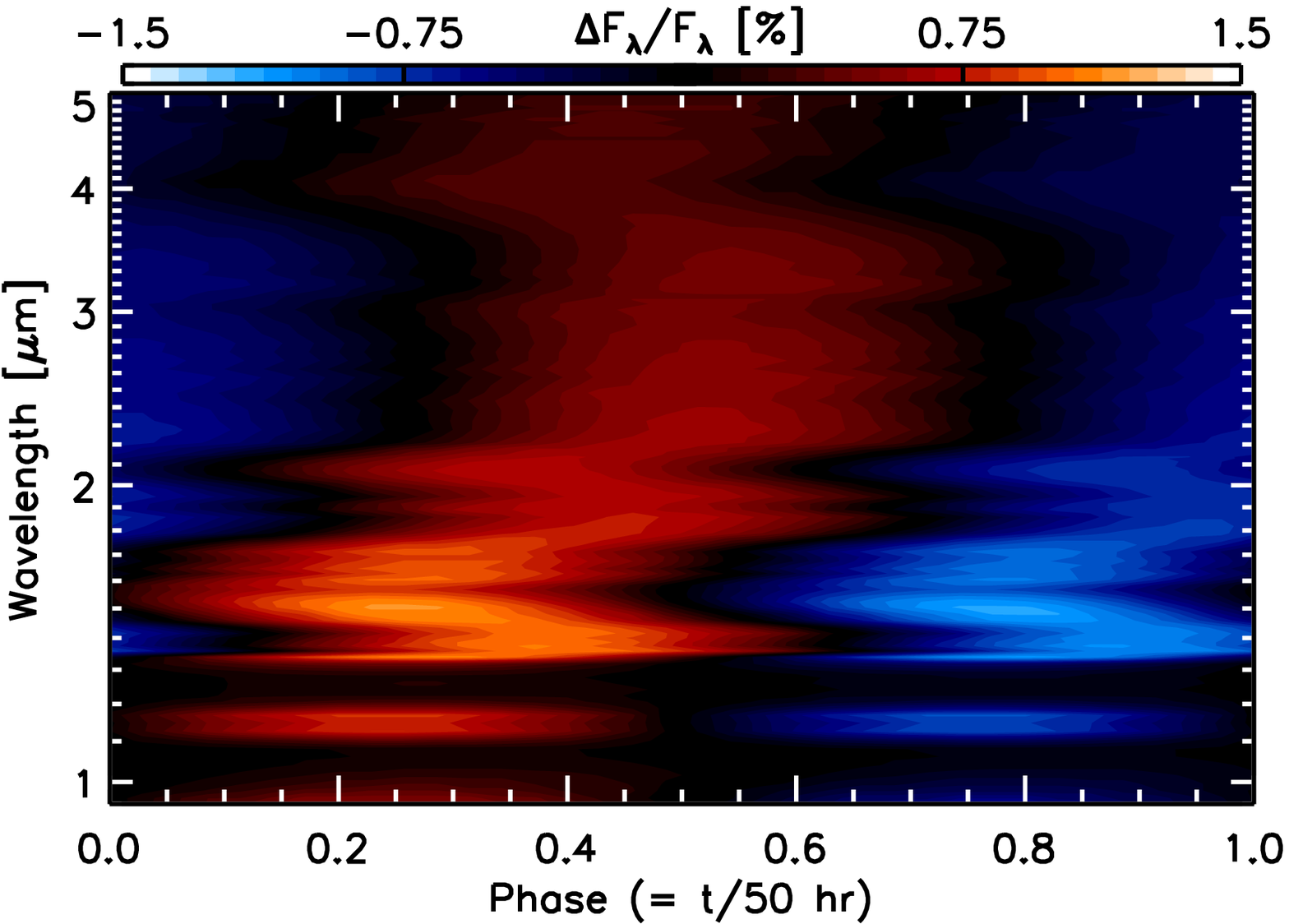}} & 
  \subfloat[]{\includegraphics[width=2.25in]{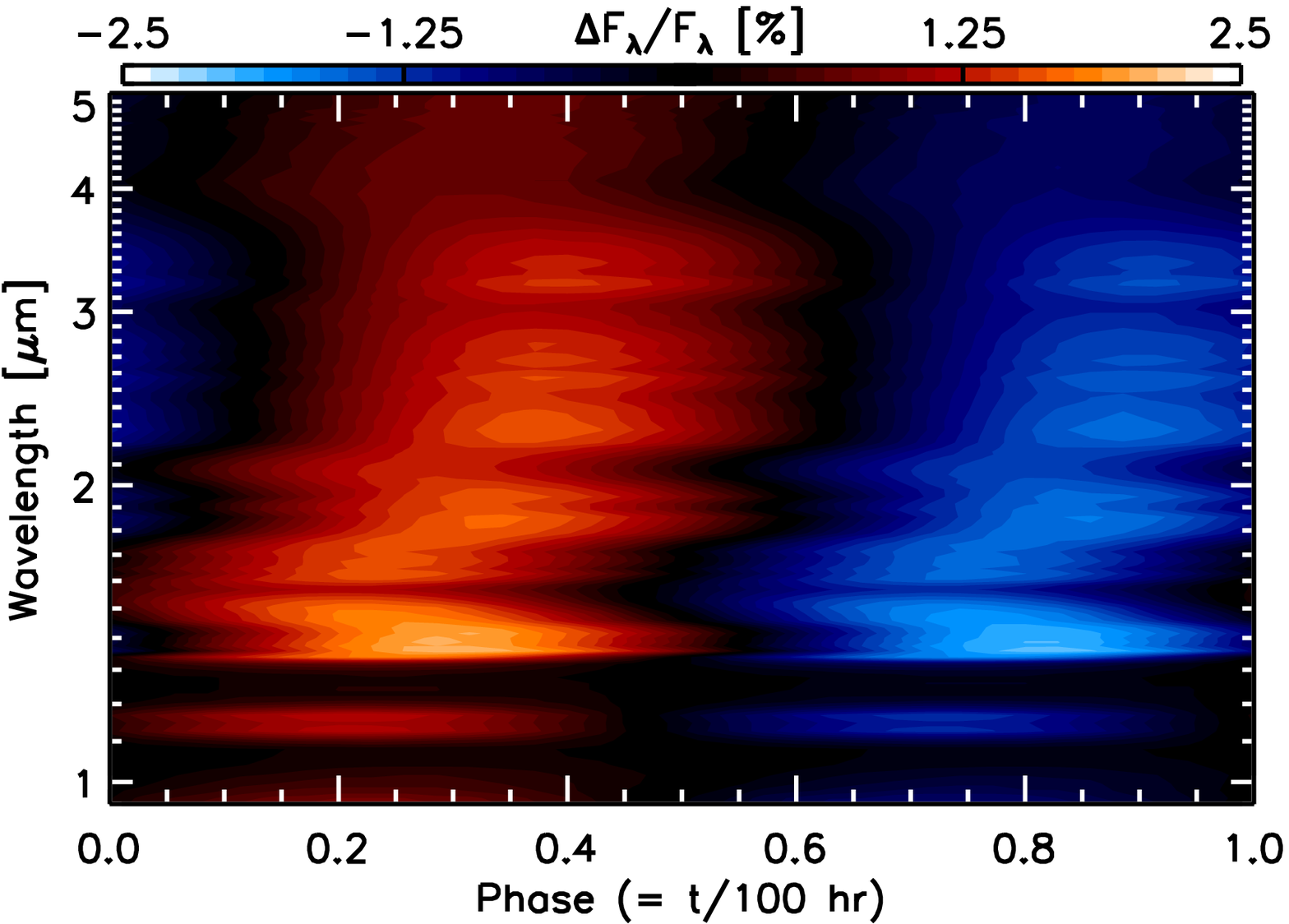}} &
  \subfloat[]{\includegraphics[width=2.25in]{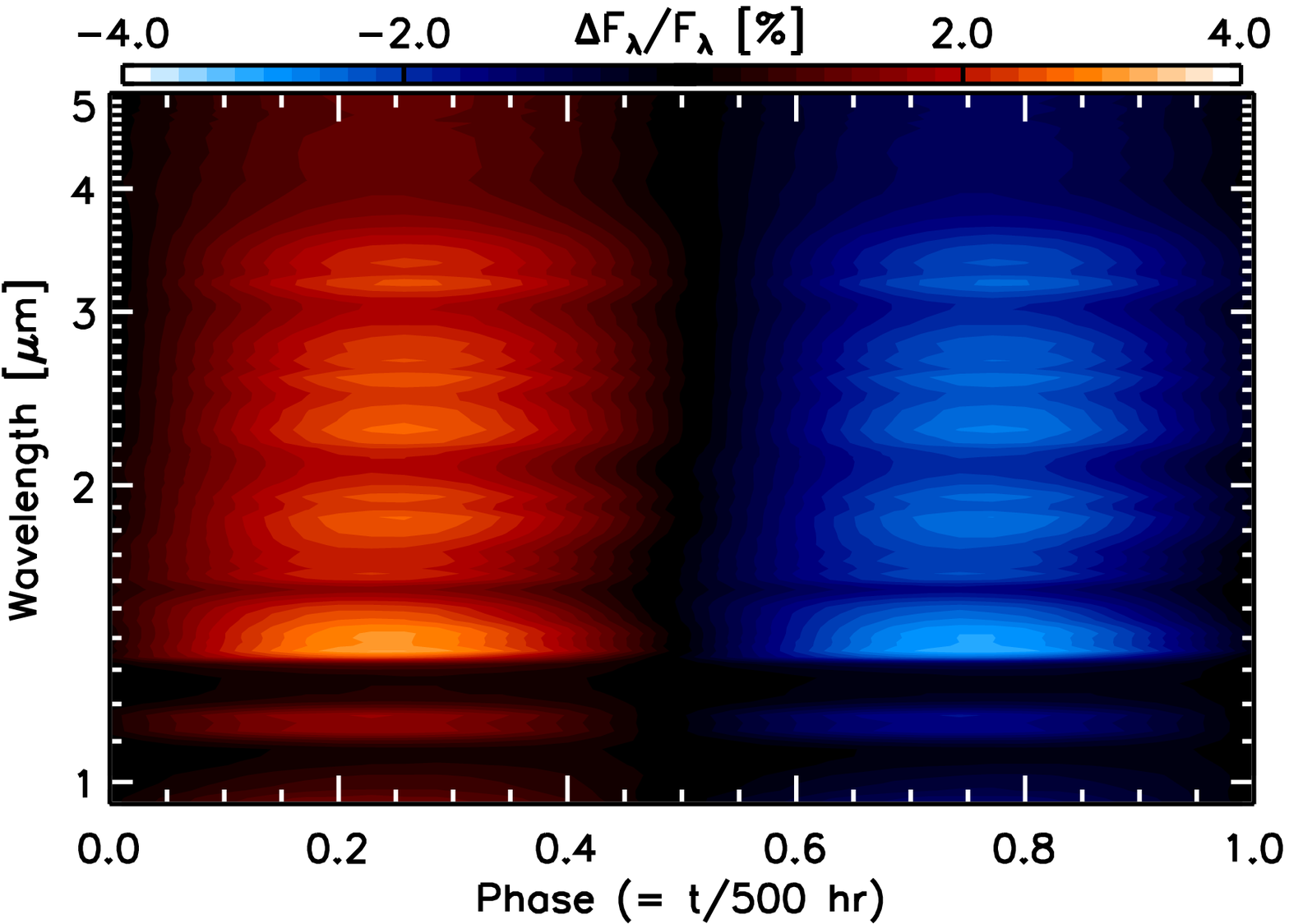}}
  \\
  \subfloat[]{\includegraphics[width=2.25in]{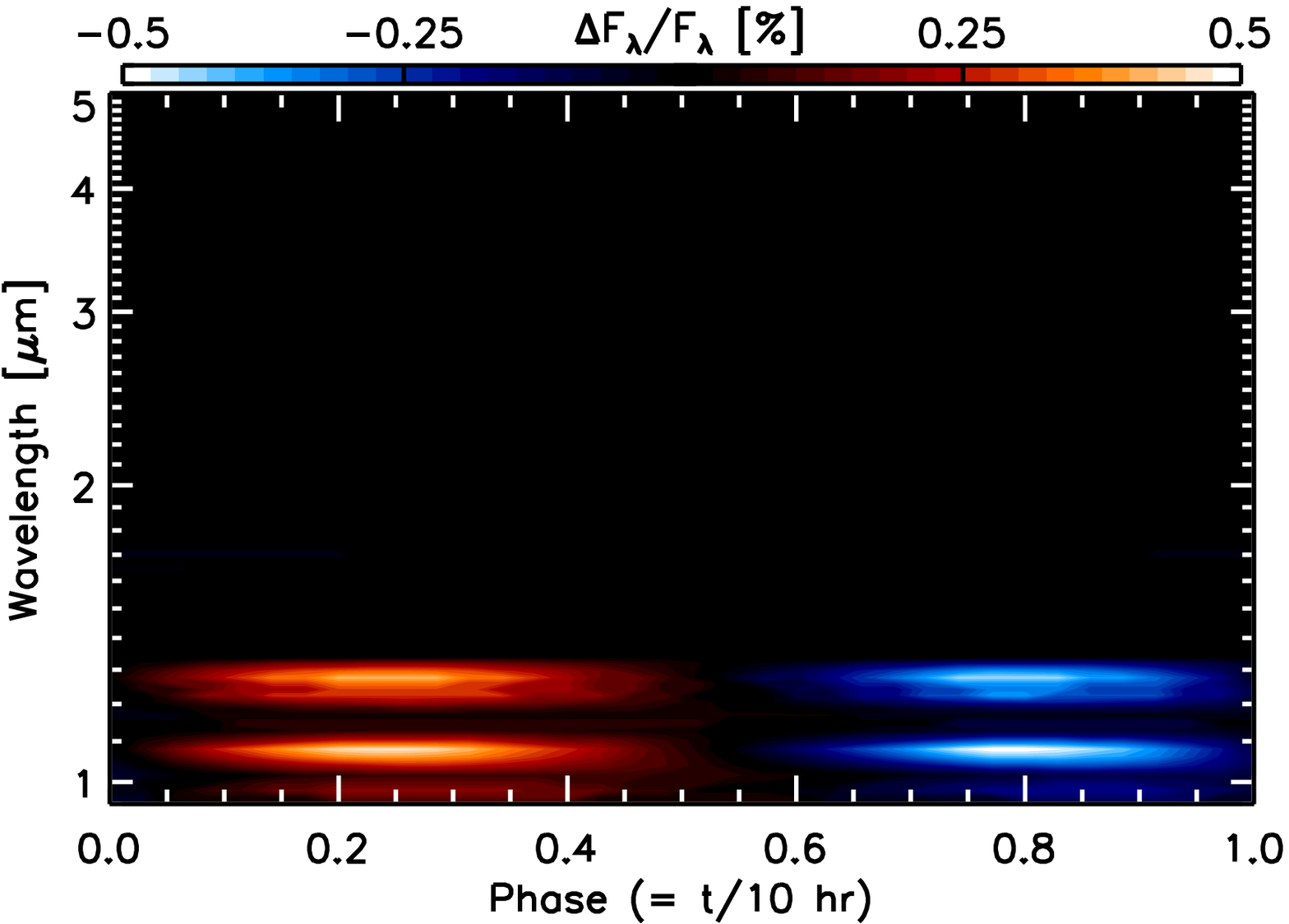}} & 
  \subfloat[]{\includegraphics[width=2.25in]{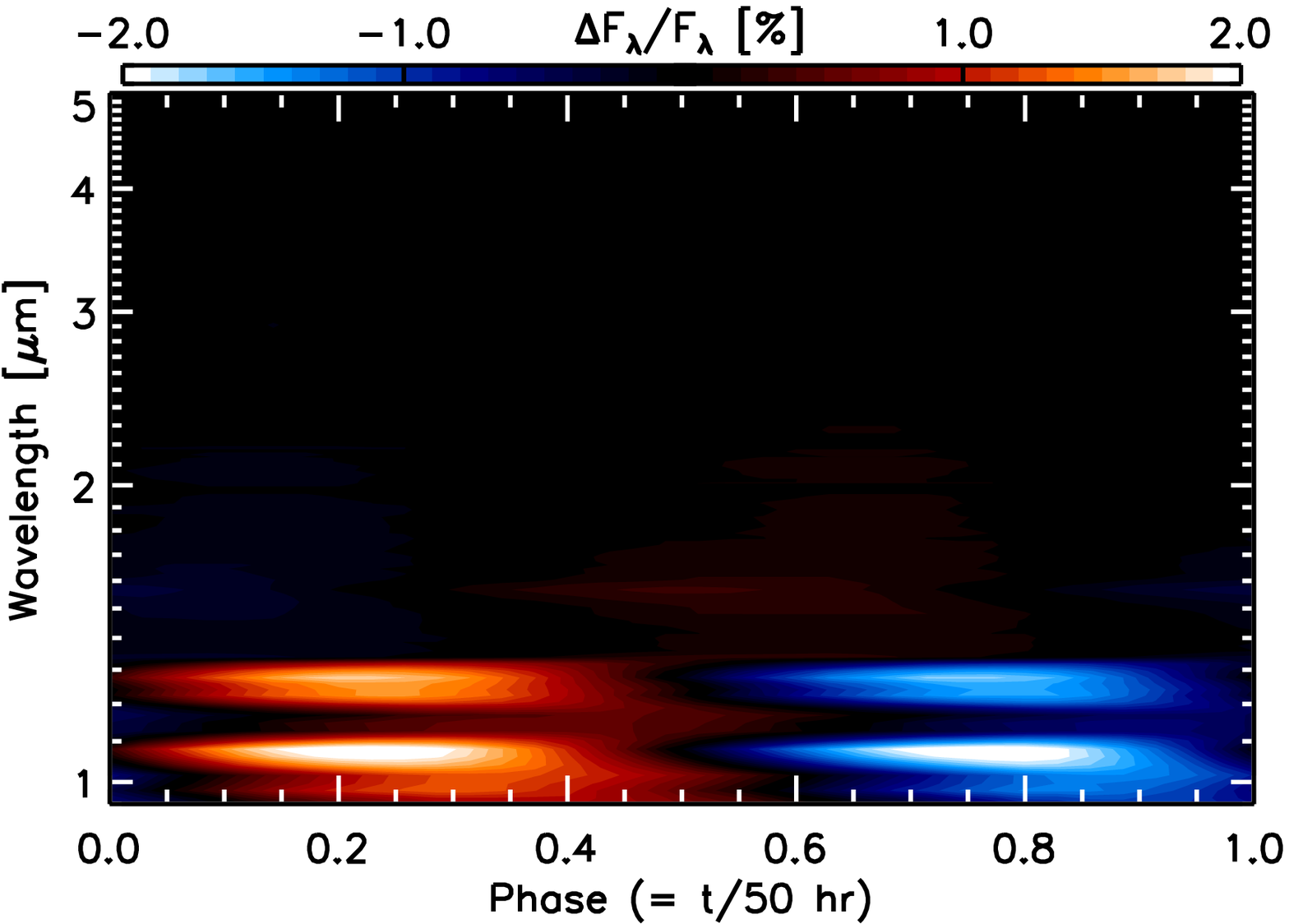}} & 
  \subfloat[]{\includegraphics[width=2.25in]{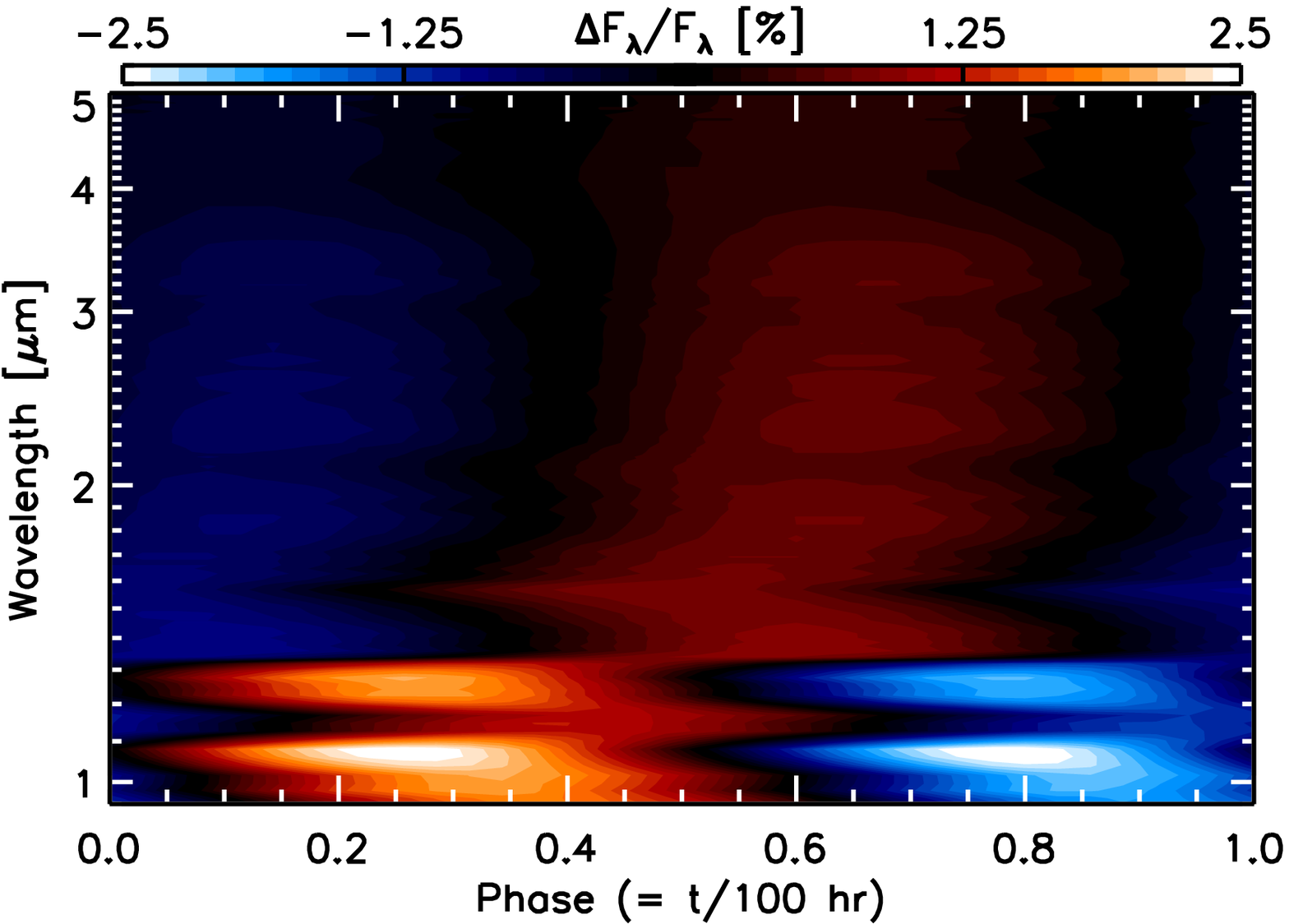}} &
  \subfloat[]{\includegraphics[width=2.25in]{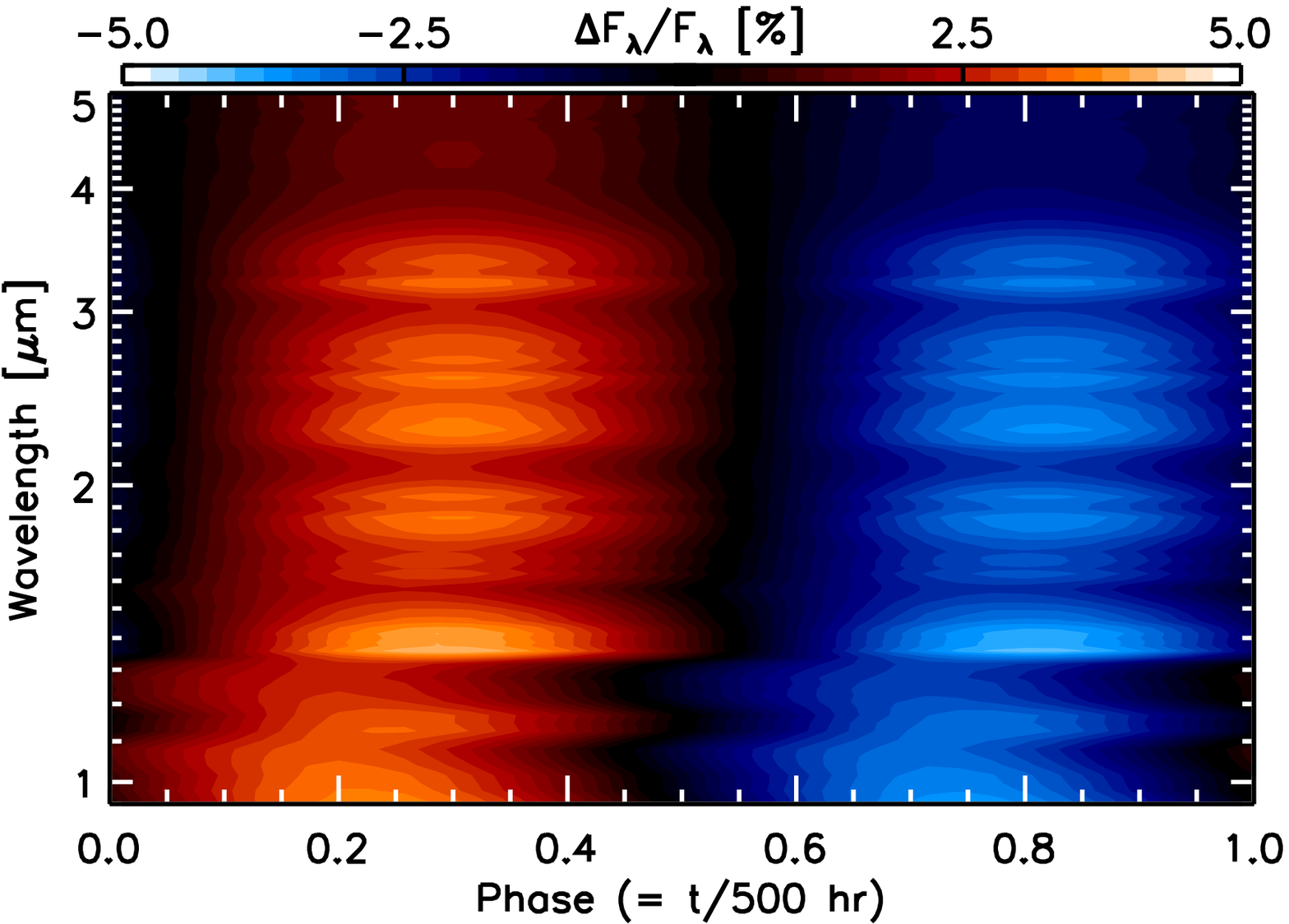}}
  \\
  \end{tabular}  
  \caption{Flux variations from time-dependent models where periodic heating perturbations 
                have been introduced at different atmosphere pressures and at different 
                timescales.  Rows are for different pressure locations of the perturbation (top - 1~bar, 
                middle - 10~bar, bottom - base of atmosphere), and columns are for different 
                timescales (left - 10~hr, mid-left - 50~hr, mid-right - 100~hr, right - 500~hr).  Horizontal 
                axis is time, which covers an entire perturbation cycle, and vertical axis is wavelength.  
                Shading indicates top-of-atmosphere brightness variations relative to their mean state 
                (i.e., averaged over an entire cycle), where red hues are brighter than average, and blue 
                hues are dimmer than average.  Note the different relative flux scales indicated at the top 
                of each sub-figure.}
  \label{fig:variability}
\end{sidewaysfigure}

\begin{figure}
  \centering
  \includegraphics[scale=0.9]{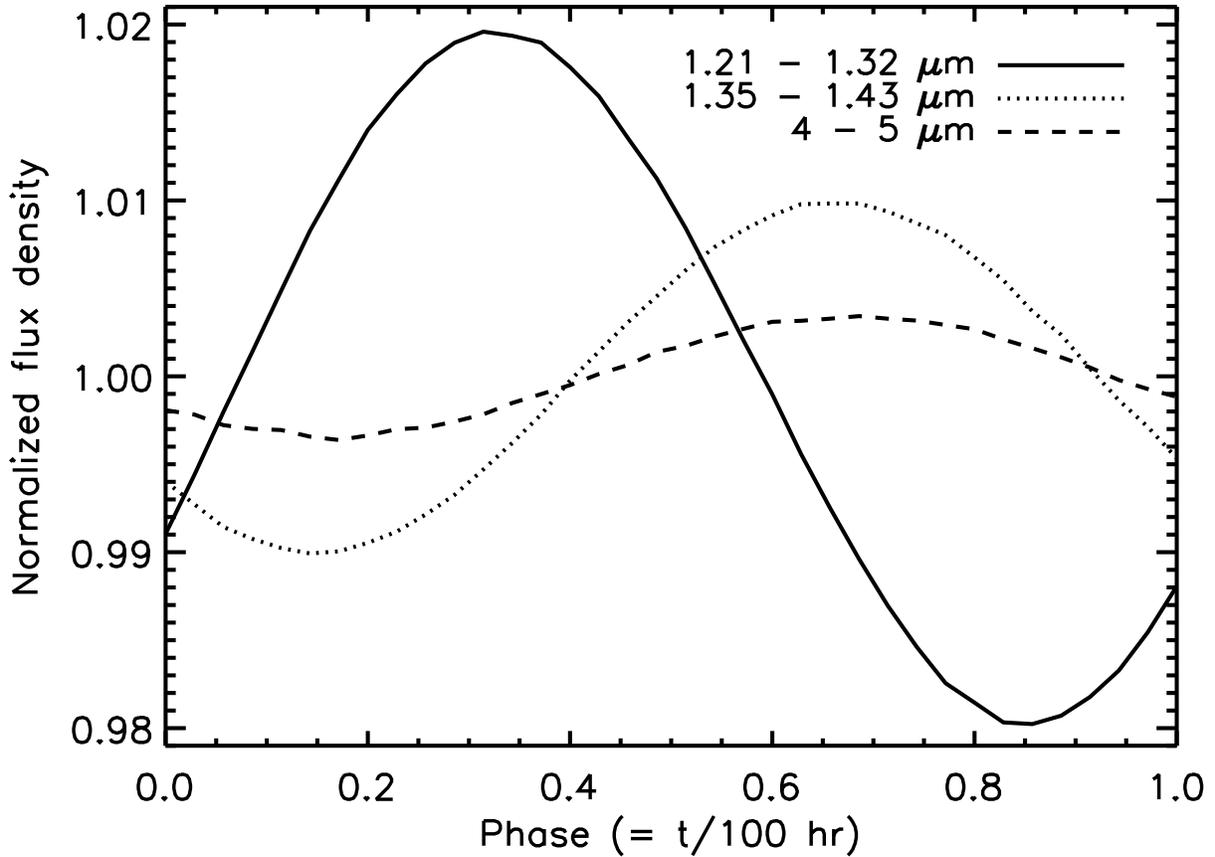}

  \caption{Example lightcurves, taken from Figure~\ref{fig:variability}, for the case where a perturbation  
                is introduced at the base of the atmosphere with a period of 100~hours.  Three different 
                bandpasses are shown (including IRAC2) that correspond to a subset of those used in 
                \citep{buenzlietal2012}, and which highlight the wavelength-dependent phase lag 
                produced by our model.}
  \label{fig:lightcurves}
\end{figure}

Figure~\ref{fig:variability} shows a grid of model results for the different perturbation 
pressures and timescales.  The horizontal axis is time, which spans one full perturbation 
cycle, and the vertical axis is wavelength through the near-infrared.  Contours show the relative 
changes in the brightness of the top-of-atmosphere spectrum ($\Delta F_{\lambda}/F_{\lambda}$, 
where $F_{\lambda}$ is the brightness averaged over a full cycle).  Recall that the 
contribution functions in Figure~\ref{fig:confuncs_spec_temp} allow us to map between 
wavelength and the pressure levels that contribute flux at that wavelength.  A set of 
lightcurves, from the case with a perturbation introduced at the base of the atmosphere 
with a period of 100~hours in Figure~\ref{fig:variability}, are shown for several different 
bandpasses in Figure~\ref{fig:lightcurves}.  These bandpasses were chosen as a 
subset of those used in \citet{buenzlietal2012}, and highlight the wavelength-dependent 
phase lag produced by our model.

\begin{figure}
  \centering
  \includegraphics[scale=0.9]{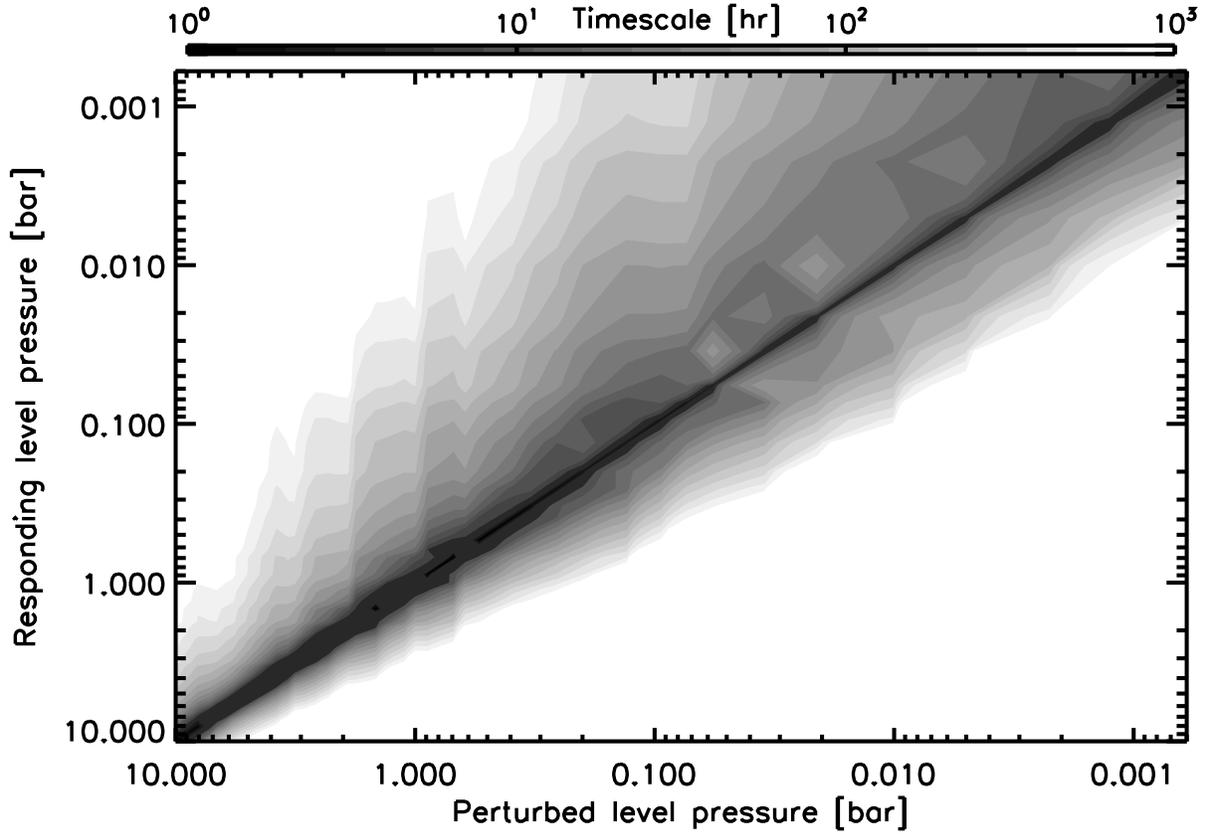}

  \caption{Atmospheric radiative response timescales.  A 1\% temperature perturbation is artificially  
                introduced at a model level (the ``perturbed level", horizontal axis), and we estimate a  
                timescale for responding to this perturbation at all other levels (the ``responding level",  
                vertical axis).  As the color bar at the top of the plot indicates, timescales range from  
                $\sim$1~hr, for levels near the perturbed level, to $\sim$$10^{3}$~hr for levels that are  
                far from the perturbation.}
  \label{fig:timescales}
\end{figure}

The models with time-varying perturbations demonstrate a number of complex 
behaviors, including wavelength-dependent phase lags.  Of note are the cases where 
the perturbations are introduced high in the atmosphere (1 bar).  Here, regardless 
of the timescale of the perturbation, the deeper atmospheric levels, which are probed 
by the shorter wavelengths, do not experience substantial temperature fluctuations, and 
so we see little or no variability at shorter wavelengths.  To understand this behavior, 
we explore the radiative response timescales in the atmosphere, which are shown in 
Figure~\ref{fig:timescales}.  We estimate the timescale at which atmospheric level $i$ 
responds to a small (1\%) temperature increase, $\Delta T$, in level $j$ by computing the 
profile of net radiative heating in response to this perturbation.  This timescale is then
\begin{equation}
\tau_{\text{rad},i} = \frac{\Delta T_{j}}{Q_{\text{rad},i}} \ ,
\end{equation}
where $Q_{rad,i}$ is the radiative heating rate at level $i$ after the perturbation is 
applied.  As Figure~\ref{fig:timescales} shows, these timescales are from $\sim$1~hr for 
levels near the perturbation, to $\sim$$10^{3}$~hr for levels far from the perturbation, 
with timescales tending to be longer for deep levels responding to perturbations at 
low pressures.

\section{Discussion}
Brown dwarf variability is influenced by a number of physical processes, each operating 
at its own timescale and spatial scale.  Commensurate with these variations will be 
thermal fluctuations, possibly due to convective motions \citep{showman&kaspi2013}, 
atmospheric wave breaking, or perhaps latent and/or cloud heating effects.  The 
model investigations presented here explore the evolution of such thermal fluctuations.

Our simplest case is an atmosphere heated from below, where the internal heat flux is 
increased, and the thermal structure is allowed to adjust.  Temperatures throughout the 
column increase by 1--2\% on timescales of order 100 hours.  As 
Figure~\ref{fig:heating_temps} demonstrates, there is clear communication between the 
deep atmosphere (at pressures larger than $\sim$10 bar) and the upper atmosphere 
(at pressures smaller than $\sim$0.1 bar).  The radiative heating rates 
(Figure~\ref{fig:heating_rates}) show that a substantial amount of heating in the upper 
atmosphere is due to radiation absorbed below about 2~$\mu$m.  The deep atmosphere, 
which is at temperatures of 1500-2000~K, contributes substantial thermal emission at 
these wavelengths (about 40\% of the flux from a 1750~K blackbody is emitted shortward of 
2~$\mu$m), and the atmosphere is relatively transparent here, too 
(Figure~\ref{fig:confuncs_spec_temp}).

Our investigation of heating from below shows that the timescale at which the atmosphere 
responds is distinct from the level radiative response timescales presented in 
Figure~\ref{fig:timescales}.  The latter simply relates a perturbation at a single level to a 
response timescale at all other atmospheric levels, whereas the former is a 
phenomenological timescale associated with the communication and passage of a 
heating (or cooling) pulse through the atmospheric column.  When we let a perturbation 
evolve in time, as when we increase the internal heat flux into the model atmosphere, we 
see that levels near the perturbed level respond rapidly (at the level radiative response 
timescale), and this communication is efficiently passed upward through the atmospheric 
column.  Thus, the level radiative response timescales tend to over-estimate the timescales 
for communicating thermal pulses over large pressure ranges.

A key feature of the radiative response times shown in Figure~\ref{fig:timescales} is 
their asymmetry in communicating a perturbation upward versus downward through the 
atmosphere.  The deep atmosphere responds more slowly to perturbations above than 
{\it vice versa}.  This is, in part, a column mass effect---an equal mass of material (per 
unit area) is located in a column extending from the top of the atmosphere to 1~bar as 
is located between 1~bar and 2~bar.  Additionally, as was mentioned above, while the 
deep atmosphere is at high enough temperatures to emit a substantial amount of 
flux at the short wavelengths where the atmosphere is relatively transparent, the upper 
atmosphere is too cool to contribute much thermal flux at these wavelengths (less 
than 1\% of the flux from a 500~K blackbody is emitted shortward of 2~$\mu$m).

The role of radiative response timescales, and their asymmetry between the deep 
atmosphere and the upper atmosphere, is apparent in fluctuations in the top-of-atmosphere 
spectrum when periodic heating perturbations are imposed at different pressures in our 
model atmosphere (Figure~\ref{fig:variability}).  When the perturbation is introduced at the 
base of the atmosphere (90~bar), and the perturbation timescale is short 
($\lesssim100$~hr), the upper atmosphere does not respond, and we see no 
variability in the spectrum between 2--5~$\mu$m (which probe lower pressures in 
the atmosphere).  As the timescale of the perturbation increases, the fluctuations 
become apparent at lower pressures.  For the longest timescale fluctuations investigated 
here (500~hr), the entire atmosphere column responds nearly in unison to the perturbation 
at depth.

When the perturbation is introduced at 10~bar, we see that the fluctuations at 
lower pressures (longer wavelengths) becomes apparent at shorter timescales.  We do 
not see fluctuations near 1.1 and 1.3~$\mu$m, which probe deeper pressures, since 
the perturbation cannot be communicated as efficiently into the deep atmosphere.  
Similar behaviors are seen when the perturbation is introduced at 1~bar, except that 
communication to pressures smaller than this occurs at shorter timescales, and so 
the fluctuations in the top-of-atmosphere spectrum look similar across a wide range of 
perturbation timescales.

The asymmetry in radiative response timescales then indicates an important 
potential observable, since spectral fluctuations due to thermal perturbations at lower 
pressures will be distinct from those occurring at depth.  Specifically, variability at 
wavelengths that probe the upper atmosphere, and a lack of variability at wavelengths 
that probe the deep atmosphere, indicates a perturbation occurring higher in the 
atmosphere.  Perturbations occurring at depth will be apparent at the wavelengths that 
probe higher pressures, and can generate spectral variability at other wavelengths, 
depending on the timescale of the perturbation.

While the extremes of the response of an atmosphere to a perturbation is uniform 
heating/cooling (at long timescales) and strictly local heating/cooling (at short 
timescales), complex behavior arises in between these extremes, and is rooted in 
the different timescales at which an atmospheric level is responding to perturbations 
in all other levels.  Wavelength-dependent phase lagging is apparent in some 
cases in Figure~\ref{fig:variability}, primarily at wavelengths below 2~$\mu$m, 
where the top-of-atmosphere spectrum probes a wide range of pressures over a 
relatively small range of wavelengths, although phase lags at longer wavelengths 
are also apparent.

We note that our case with a perturbation introduced at the base of the atmosphere 
on a timescale of 100 hours produces brightness fluctuations whose wavelength 
dependence bear a similarity to those observed in \citet{buenzlietal2012} (see 
Figure~\ref{fig:lightcurves}).  Like the observations reported by Buenzli~et~al., the 
amplitude of the model fluctuations is typically of order 1--3\%, and the flux measured in 
a 1.35--1.43~$\mu$m bandpass lags the flux measured in a 1.21--1.32~$\mu$m bandpass 
(by 140$^{\circ}$ in our model, as compared to $\sim$180$^{\circ}$ in the observations).  
However, our model cannot explain all of the details of the observations.  Most importantly, 
the brown dwarf varies on roughly hour-long timescales, which is comparable
to the rotation period and probably reflects the complex, three-dimensional nature of the 
atmosphere.  In contrast, our model perturbation timescales are closer to 100~hours.  This is 
likely a shortcoming of using a one-dimensional model, although we note that the observations 
reported by \citet{buenzlietal2012} do show brightness trends on timescales longer than 
10~hours.  Thus, in summary, our results show that thermal fluctuations can be an 
important aspect of spectral variability, while also pointing to the necessary role of 
dynamics and clouds in fully explaining observations.

Future studies should examine the interconnected role of thermal perturbations and cloud
and chemical evolution on thermal variability.  Clouds typically act as a continuum opacity 
source.  Thus, clouds will limit sensitivity to thermal fluctuations occurring in atmospheric 
layers beneath the cloud base (i.e., those layers probed at window/continuum wavelengths 
in the cloud-free case), and this effect will be stronger for more optically thick clouds.  
Alternatively, clouds will shield deeper atmospheric layers from thermal perturbations that are 
propagating downwards, preventing those layers from heating/cooling as they would in a 
cloud-free scenario.  Of course, clouds are not a steady-state process, and so will respond 
with their own characteristic timescales when other atmospheric layers are perturbed from 
their equilibrium state.  For example, if an amount of flux, $\Delta F$, across a cloud with 
column mass $M_{c}$ (analogous to liquid water path for Earth clouds) goes into vaporizing 
the cloud particles, then a characteristic timescale would be 
$\tau_{\text{cld}} \sim M_{c}L/\Delta F$, where $L$ is the latent heat of vaporization.  Thus, 
there is the potential for interesting feedbacks to occur, where the atmosphere is attempting 
to respond at one timescale (like those discussed in this work), while the clouds are evolving 
at another timescale.

Finally, as the ability of different layers in the atmosphere to influence one another via radiation 
hinges on the relatively clear opacity spectral windows found in T-dwarf atmospheres.  L-dwarfs, 
with stronger cloud and molecular continuum opacity which partially fill in the spectral windows, 
may accordingly exhibit less radiative coupling between layers, longer response timescales, and 
perhaps smaller phase shifts between different spectral bandpasses. Larger systematic surveys 
into brown dwarf variability should investigate such possibilities, not to mention the 
three-dimensional roles of atmospheric dynamics as coupled to rotation.

\section{Conclusions}
Using a one-dimensional model of brown dwarf atmospheric structure, we have studied 
the time-dependent evolution of the atmosphere in response to a variety of thermal 
perturbations.  We omitted cloud and dynamical effects, choosing to concentrate on 
behaviors that arise strictly due to atmospheric thermal variations.  Thermal perturbations 
of the deep atmosphere can be communicated to the upper atmosphere at shorter 
near-infrared wavelengths, although communication in the opposite direction is impeded 
by the lack of flux generated at these wavelengths by the relatively cool upper atmosphere.  
The response timescale of the atmosphere to thermal perturbations is typically 
10--100~hours.  Deep thermal perturbations can lead to brightness fluctuations at nearly all 
near-infrared wavelengths, and our model predicts that these could be observed on timescales 
of hundreds of hours.  While it is not our goal to solve the entire problem of brown dwarf variability, 
our model can produce a number of the observed features, depending on the nature of 
the thermal perturbation.  In the future, a full explanation of variability in brown dwarf thermal 
emission spectra must incorporate three-dimensional atmospheric and cloud dynamics, as 
well as the time-dependent evolution of thermal perturbations throughout the radiative portion 
of the atmosphere.

\acknowledgements
TR gratefully acknowledges support from an appointment to the NASA 
Postdoctoral Program at NASA Ames Research Center, administered by 
Oak Ridge Affiliated Universities.  MM acknowledges support of the NASA Planetary
Atmospheres and Origins programs. We thank Jonathan Fortney for sharing 
tools for computing normalized contribution functions, Jacqueline Radigan 
for reviewing an early version of this paper, and Caroline Morley for sharing updates 
to relevant cloud routines in our thermal structure model.
%


\begin{thebibliography}{}
\expandafter\ifx\csname natexlab\endcsname\relax\def\natexlab#1{#1}\fi

\bibitem[{{Ackerman} \& {Marley}(2001)}]{ackerman&marley2001}
{Ackerman}, A.~S., \& {Marley}, M.~S. 2001, \apj, 556, 872

\bibitem[{{Allard} {et~al.}(2001){Allard}, {Hauschildt}, {Alexander},
  {Tamanai}, \& {Schweitzer}}]{allardetal2001}
{Allard}, F., {Hauschildt}, P.~H., {Alexander}, D.~R., {Tamanai}, A., \&
  {Schweitzer}, A. 2001, \apj, 556, 357

\bibitem[{{Apai} {et~al.}(2013){Apai}, {Radigan}, {Buenzli}, {Burrows}, {Reid},
  \& {Jayawardhana}}]{apaietal2013}
{Apai}, D., {Radigan}, J., {Buenzli}, E., {et~al.} 2013, \apj, 768, 121

\bibitem[{{Artigau} {et~al.}(2009){Artigau}, {Bouchard}, {Doyon}, \&
  {Lafreni{\`e}re}}]{artigauetal2009}
{Artigau}, {\'E}., {Bouchard}, S., {Doyon}, R., \& {Lafreni{\`e}re}, D. 2009,
  \apj, 701, 1534

\bibitem[{{Bailer-Jones} \& {Mundt}(1999)}]{bailerjones&mundt1999}
{Bailer-Jones}, C.~A.~L., \& {Mundt}, R. 1999, \aap, 348, 800

\bibitem[{{Bailer-Jones} \& {Mundt}(2001)}]{bailerjones&mundt2001a}
---. 2001, \aap, 367, 218

\bibitem[{{Buenzli} {et~al.}(2013){Buenzli}, {Apai}, {Radigan}, {Reid}, \&
  {Flateau}}]{buenzlietal2013}
{Buenzli}, E., {Apai}, D., {Radigan}, J., {Reid}, I.~N., \& {Flateau}, D. 2013,
  ArXiv e-prints, arXiv:1312.5294

\bibitem[{{Buenzli} {et~al.}(2012){Buenzli}, {Apai}, {Morley}, {Flateau},
  {Showman}, {Burrows}, {Marley}, {Lewis}, \& {Reid}}]{buenzlietal2012}
{Buenzli}, E., {Apai}, D., {Morley}, C.~V., {et~al.} 2012, \apjl, 760, L31

\bibitem[{{Burgasser} {et~al.}(2004){Burgasser}, {McElwain}, {Kirkpatrick},
  {Cruz}, {Tinney}, \& {Reid}}]{burgasseretal2004}
{Burgasser}, A.~J., {McElwain}, M.~W., {Kirkpatrick}, J.~D., {et~al.} 2004,
  \aj, 127, 2856

\bibitem[{{Burrows} {et~al.}(2006){Burrows}, {Sudarsky}, \&
  {Hubeny}}]{burrowsetal2006}
{Burrows}, A., {Sudarsky}, D., \& {Hubeny}, I. 2006, \apj, 640, 1063

\bibitem[{{Chabrier} {et~al.}(2000){Chabrier}, {Baraffe}, {Allard}, \&
  {Hauschildt}}]{chabrieretal2000}
{Chabrier}, G., {Baraffe}, I., {Allard}, F., \& {Hauschildt}, P. 2000, \apj,
  542, 464

\bibitem[{{Freedman} {et~al.}(2008){Freedman}, {Marley}, \&
  {Lodders}}]{freedmanetal2008}
{Freedman}, R.~S., {Marley}, M.~S., \& {Lodders}, K. 2008, \apjs, 174, 504

\bibitem[{{Freytag} {et~al.}(2010){Freytag}, {Allard}, {Ludwig}, {Homeier}, \&
  {Steffen}}]{freytagetal2010}
{Freytag}, B., {Allard}, F., {Ludwig}, H.-G., {Homeier}, D., \& {Steffen}, M.
  2010, \aap, 513, A19

\bibitem[{{Gelino} \& {Marley}(2000)}]{gelino&marley2000}
{Gelino}, C., \& {Marley}, M. 2000, in Astronomical Society of the Pacific
  Conference Series, Vol. 212, From Giant Planets to Cool Stars, ed. C.~A.
  {Griffith} \& M.~S. {Marley}, 322

\bibitem[{{Gelino} {et~al.}(2002){Gelino}, {Marley}, {Holtzman}, {Ackerman}, \&
  {Lodders}}]{gelinoetal2002}
{Gelino}, C.~R., {Marley}, M.~S., {Holtzman}, J.~A., {Ackerman}, A.~S., \&
  {Lodders}, K. 2002, \apj, 577, 433

\bibitem[{{Gierasch} \& {Goody}(1968)}]{gierasch&goody1968}
{Gierasch}, P., \& {Goody}, R. 1968, \planss, 16, 615

\bibitem[{{Gillon} {et~al.}(2013)}]{gillonetal2013} {Gillon}, M., {Triaud}, 
A.~H.~M.~J., {Jehin}, E., {et~al.} 2013, \aap, 555, L5 

\bibitem[{{Golimowski} {et~al.}(2004){Golimowski}, {Leggett}, {Marley}, {Fan},
  {Geballe}, {Knapp}, {Vrba}, {Henden}, {Luginbuhl}, {Guetter}, {Munn},
  {Canzian}, {Zheng}, {Tsvetanov}, {Chiu}, {Glazebrook}, {Hoversten},
  {Schneider}, \& {Brinkmann}}]{golimowskietal2004}
{Golimowski}, D.~A., {Leggett}, S.~K., {Marley}, M.~S., {et~al.} 2004, \aj,
  127, 3516

\bibitem[{{Hanel} {et~al.}(1992){Hanel}, {Conrath}, {Jennings}, \&
  {Samuelson}}]{haneletal1992}
{Hanel}, R.~A., {Conrath}, B.~J., {Jennings}, D.~E., \& {Samuelson}, R.~E.
  1992, {Exploration of the solar system by infrared remote sensing} (Cambridge
  University Press)

\bibitem[{{Knapp} {et~al.}(2004){Knapp}, {Leggett}, {Fan}, {Marley}, {Geballe},
  {Golimowski}, {Finkbeiner}, {Gunn}, {Hennawi}, {Ivezi{\'c}}, {Lupton},
  {Schlegel}, {Strauss}, {Tsvetanov}, {Chiu}, {Hoversten}, {Glazebrook},
  {Zheng}, {Hendrickson}, {Williams}, {Uomoto}, {Vrba}, {Henden}, {Luginbuhl},
  {Guetter}, {Munn}, {Canzian}, {Schneider}, \& {Brinkmann}}]{knappetal2004}
{Knapp}, G.~R., {Leggett}, S.~K., {Fan}, X., {et~al.} 2004, \aj, 127, 3553

\bibitem[{{Leggett} {et~al.}(1998){Leggett}, {Allard}, \&
  {Hauschildt}}]{leggettetal1998}
{Leggett}, S.~K., {Allard}, F., \& {Hauschildt}, P.~H. 1998, \apj, 509, 836

\bibitem[{Lodders(2004)}]{lodders2004}
Lodders, K. 2004, Science, 303, 323

\bibitem[{{Lodders} \& {Fegley}(2006)}]{lodders&fegley2006}
{Lodders}, K., \& {Fegley}, Jr., B. 2006, in Astrophysics Update 2, ed. J.~W.
  {Mason} ({Springer-Praxis Books}), 1--28

\bibitem[{{Marley} {et~al.}(1996){Marley}, {Saumon}, {Guillot}, {Freedman},
  {Hubbard}, {Burrows}, \& {Lunine}}]{marleyetal1996}
{Marley}, M.~S., {Saumon}, D., {Guillot}, T., {et~al.} 1996, Science, 272, 1919

\bibitem[{{Marley} {et~al.}(2002){Marley}, {Seager}, {Saumon}, {Lodders},
  {Ackerman}, {Freedman}, \& {Fan}}]{marleyetal2002}
{Marley}, M.~S., {Seager}, S., {Saumon}, D., {et~al.} 2002, \apj, 568, 335

\bibitem[{Morley {et~al.}(2012)Morley, Fortney, Marley, Visscher, Saumon, \&
  Leggett}]{morleyetal2012}
Morley, C.~V., Fortney, J.~J., Marley, M.~S., {et~al.} 2012, The Astrophysical
  Journal, 756, 172

\bibitem[{{Radigan} {et~al.}(2012){Radigan}, {Jayawardhana}, {Lafreni{\`e}re},
  {Artigau}, {Marley}, \& {Saumon}}]{radiganetal2012}
{Radigan}, J., {Jayawardhana}, R., {Lafreni{\`e}re}, D., {et~al.} 2012, \apj,
  750, 105

\bibitem[{{Saumon} {et~al.}(1995){Saumon}, {Chabrier}, \& {van
  Horn}}]{saumonetal1995}
{Saumon}, D., {Chabrier}, G., \& {van Horn}, H.~M. 1995, \apjs, 99, 713

\bibitem[{{Showman} \& {Kaspi}(2013)}]{showman&kaspi2013}
{Showman}, A.~P., \& {Kaspi}, Y. 2013, \apj, 776, 85

\bibitem[{{Stephens} {et~al.}(2009){Stephens}, {Leggett}, {Cushing}, {Marley},
  {Saumon}, {Geballe}, {Golimowski}, {Fan}, \& {Noll}}]{stephensetal2009}
{Stephens}, D.~C., {Leggett}, S.~K., {Cushing}, M.~C., {et~al.} 2009, \apj,
  702, 154

\bibitem[{{Tinney} \& {Tolley}(1999)}]{tinney&trolley1999}
{Tinney}, C.~G., \& {Tolley}, A.~J. 1999, \mnras, 304, 119

\bibitem[{Toon {et~al.}(1989)Toon, McKay, Ackerman, \&
  Santhanam}]{toonetal1989}
Toon, O.~B., McKay, C., Ackerman, T., \& Santhanam, K. 1989, Journal of
  Geophysical Research: Atmospheres (1984--2012), 94, 16287

\bibitem[{{Tsuji}(2002)}]{tsuji2002}
{Tsuji}, T. 2002, \apj, 575, 264

\bibitem[{{Vitense}(1953)}]{vitense1953}
{Vitense}, E. 1953, \zap, 32, 135

\bibitem[{{Young} {et~al.}(1997){Young}, {Yelle}, {Young}, {Seiff}, \&
  {Kirk}}]{youngetal1997}
{Young}, L.~A., {Yelle}, R.~V., {Young}, R., {Seiff}, A., \& {Kirk}, D.~B.
  1997, Science, 276, 108

\end{thebibliography}




%
%

%
\end{document}